\documentclass[useAMS,usenatbib]{mn2e}

\usepackage{graphicx}

\begin{document}
\title{Applications  of Integrated Photonic Spectrographs in Astronomy}


\author[R. J. Harris and J. R. Allington-Smith]{R.J. Harris$^{1}$\thanks{E-mail:
r.j.harris@durham.ac.uk (RJH)} and J. R. Allington-Smith$^{1}$ \\
$^{1}$ Centre for Advanced Instrumentation, Physics Dept, Durham University, South Rd, Durham, DH1 3LE}

\maketitle
\begin{abstract}

One of the  problems of producing instruments for Extremely Large Telescopes is that their size (and hence cost) scales rapidly with telescope aperture. To try to break this relation alternative new technologies have been proposed, such as the use of the Integrated Photonic Spectrograph (IPS). Due to their diffraction-limited nature the IPS is claimed to defeat the harsh scaling law applying to conventional instruments. In contrast to  photonic applications, devices for astronomy are not usually used at the diffraction limit. Therefore to retain throughput and spatial information, the IPS requires a photonic lantern (PL)  to decompose the input multimode light into single modes. This is then fed into either numerous Arrayed Waveguide Gratings (AWGs) or a conventional spectrograph. \newline
 We investigate the potential advantage of using an IPS instead of conventional monolithic optics for a variety of capabilities represented by existing instruments on 8m telescopes and others planned for Extremely Large Telescopes (ELTs). To do this, we have constructed toy models of different versions of the IPS and calculated the relative instrument sizes and the number of detector pixels required. This allows us to quantify the relative size/cost advantage for instruments aimed at different science requirements.
We show that a full IPS instrument is equivalent to an image-slicer. Image-slicing is a beneficial strategy for ELTs as previously demonstrated. However, the requirement to decompose the input light into individual modes imposes a redundancy in terms of the numbers of components and detector pixels in many cases which acts to cancel out the advantage of the small size of the photonic components. However, there are specific applications where an IPS gives a potential advantage which we describe. Furthermore, the IPS approach has the potential advantage of minimising or eliminating bulk optics.  We show that AWGs fed with multiple single-mode inputs from a PL require relatively bulky auxiliary optics and a 2-D detector array which significantly increases the size of the instrument. A more attractive option is  to combine the outputs of many AWGs so that a 1-D detector can be used to greatly  reduce the number of detector pixels required and  provide 
efficient adaptation to the curved output focal surface.

\end{abstract}


\section{Introduction}

Spectroscopy is one of the most useful tools in astronomy, with applications in fields ranging from Cosmology to Exoplanet studies. Arguably the most common form in astronomy is dispersive spectroscopy which uses a dispersive element to separate different wavelengths. \newline
In its simplest form a dispersive spectrograph contains four components: A slit to isolate the area to be dispersed, a collimator, a grating or prism to disperse the light and a camera and detector to record the intensity at each wavelength. In order to retain throughput the slit of the spectrograph is usually matched to the seeing of the telescope. If this is not diffraction-limited then the collimated beam must increase in size as the telescope grows in size in order to maintain the same resolution \citep{Lee2000}. The immediate consequence of this  is that the optics and disperser must also grow in size. This then leads physical problems such as stresses and flexure in the materials, along with the difficulties inherent in building these large monolithic structures. \newline
In order to use the same design principles as existing instruments more exotic materials and construction processes are needed, which drives the costs of building the instruments up. Using the conventional approach the cost of instruments scales with at least the square of the telescope aperture \citep{Bland-Hawthorn2006}. \newline
In order to reduce these problems the input to the spectrograph can be sliced by dividing the input field (the slit) into a number of thin slices. Each of these can then  be fed into a spectrograph or spectrographs. This technique is known as image slicing. Each of these spectrographs produces a spectrum for each slice. The resulting data is then reformatted into a 3D datacube with axes x,y,$\lambda$ allowing the reconstruction of the initial image. \newline
There are several different methods for image slicing detailed in the literature, from lenslet arrays feeding fibres to slicing mirrors \citep{Allington-Smith}. All have inherent advantages and disadvantages. The various methods have all been widely adopted in recent years and are used in various forms in instruments on the current generation of 8m telescopes. Theoretical investigations into the scaling laws in relation to image slicing suggest it will be an even more powerful tool to reduce instrument sizes and costs on the next generation of 30m telescopes \citep{Allington-Smith2009a}. \newline

In this paper we examine the potential applications of the Integrated Photonic Spectrograph (IPS), which shares some of the features of image slicers. Despite the physical differences they obey the same basic Physical laws as conventional instruments. \newline
The IPS devices take light from an input  fibre which is usually matched to the seeing limit (as with conventional fibre fed instruments) and so supports many modes. The light from this multimode fibre (MMF) is then split into a number of single mode fibres (SMF) by a photonic lantern (PL). At this point two options have been proposed \citep{Bland-Hawthorn2010}. \newline

\noindent \textbf{P=1:} The first requires a reformatting component \citep{2012OExpr..2013996B, Thomson2011} to form a slit of SMFs which can then be dispersed by bulk optics.  We shall call this the semi-photonic case. \newline
\textbf{P=2:} In the second the SMFs are then fed into Arrayed Waveguide Gratings (AWGs) which disperse the light into individual spectra, with one or more specta per AWG \citep{Bland-Hawthorn2010}. These spectra have the advantage that they are in a linear format, so can be sampled using an array of fibres or a linear detector. This is our fully-photonic case. \newline

While the principles for both have been demonstrated \citep{2012arXiv1208.3006L,Cvetojevic2009,Cvetojevic2012}, 
so far these have only used single or few modes from a single input, resulting in limitations in throughput and field. They are also not optimised in terms of size or the components used. 
This led us to investigate  how complete instruments would perform compared to current instrumentation, with initial results suggesting that a fully-photonic  IPS would be best suited to small diffraction-limited telescopes with small fields of view \citep{Harris2012}. 

In this paper we determine the application areas where the IPS may have an advantage over conventional instrumentation.
After noting the formal similarity between image-slicing and photonic spectroscopy in section \ref{sec:modes},  we  consider the requirement for the field of view of the instrument in section \ref{sec:fov}.  Simplified models of the IPS are presented in sections \ref{sec:echelle} and \ref{sec:awg}. The results of comparing  conventional and IPS instruments is given in section \ref{sec:results}. In section \ref{sec:cross_disp}, we discuss different ways to reduce the number of AWGs and/or detector pixels, before presenting our conclusions in section \ref{sec:conclusion}.

\section{Instrument size scale relationships}
\label{sec:modes}

It is often claimed that IPS violates the relationship between telescope diameter and spectral resolution:

\begin{equation}
	R =  \frac{m \rho \lambda W}{\chi D_{T}}  = \frac{2 \tan \gamma D_{col}}{\chi D_{T}} 
	\label{eqn:res}
\end{equation}
where \textit{R} is the resolution of the instrument, \textit{m} is the diffraction order, $\rho$ is the ruling density, $\lambda$ is the wavelength, \textit{W} is the length of intersection between the grating and collimated beam, $\chi$ is the angular slitwidth, $D_{T}$ is the diameter of the telescope, $\gamma$ is the blaze angle, $\rho$ is the ruling density and $D_{col}$ the diameter of the collimated beam.

This applies to a slit spectrograph using a diffraction grating as the dispersive element and shows that for a given resolution, angular slitwidth and blaze angle that the diameter of the collimated beam must increase in proportion to the telescope aperture, leading to a bigger instrument. 

Unlike a conventional spectrograph the input to the IPS must be diffraction-limited ($\lambda \approx \chi D_{T}$) due to its single mode nature, so the resolution can be shown to be of the form 

\begin{equation}
	R  = \frac{mN_{wg}}{C}.
	\label{eqn:awg_res}
\end{equation}

where \textit{$N_{wg}$} is the number of waveguides in the AWG model or number of rulings on a conventional grating and \textit{C} a factor to account for manufacturing errors  \citep{Lawrence2010}. This has no dependence on telescope diameter so it would appear to break the relation. It must be noted though, this applies to a device operated at the diffraction limit of the telescope, not at the seeing limit as with equation \ref{eqn:res}. 

To examine what happens when the input at the seeing limit we consider the number of spatial modes in a conventional step-index fibre \citep{cheo} which can be approximated as

\begin{equation}
	M = \frac{V_{fibre}^{2}}{4} .
\end{equation}

It is useful to remember that each spatial mode has two polarisation states, though we do not include the factor here as each single mode fibre accepts two polarisations. The associated $V$ parameter is 

\begin{equation}
	V_{\rm fibre} = \frac{ \pi s \Theta}{\lambda} .
\end{equation}
Where \textit{s} is the diameter of the fibre core (assumed equal  to the slitwidth in equation \ref{eqn:res}) and $\Theta$ the numerical aperture at which the fibre is operated. This must be less than the limiting (i.e. maximum) numerical aperture of the fibre. Noting that
\begin{eqnarray}
	\Theta &\approx& \frac{1}{2 F_{T}} \\
	s &=& \chi f_{T} = \chi F_{T} D_{T} 
\end{eqnarray} 

\noindent where $F_{T}$ the telescope focal ratio and $f_{T}$ is the telescope focal length, the number of modes is given by

\begin{equation}
	M =  \left( \frac{\pi \chi D_{T}}{4 \lambda} \right)^{2}.
	\label{eqn:modes}
\end{equation}

Therefore it can be seen that for each sampling element the number of modes increases as the square of the telescope diameter in a similar way to the number of slices at the diffraction limit $(\chi D_{T} / 1.22 \lambda )^{2}$. 
This confirms that, to first order, photonic spectrographs are bound by the same scaling laws as conventional spectrographs. In what follows, we attempt to quantify areas where photonic spectrographs may confer an advantage, and suggest modifications which may allow photonic spectrographs to make a significant impact on future astronomical instrumentation.

\section{The input field and spatial multiplex}
\label{sec:fov}

In the previous section we calculated the number of modes per spatial sampling element (spaxel). As with diverse field spectroscopy \citep{Murray2009}, photonic spectrographs address a number of individual spaxels, which can be grouped (as in Integral Field Spectroscopy; IFS), or separate (as in Multi Object Spectroscopy;  MOS).  In order to fairly compare with conventional instrumentation we need to make sure we sample the same number of spaxels (e.g. observe the same field). 

\begin{figure*}
	\begin{center}
		\includegraphics[width = 168mm]{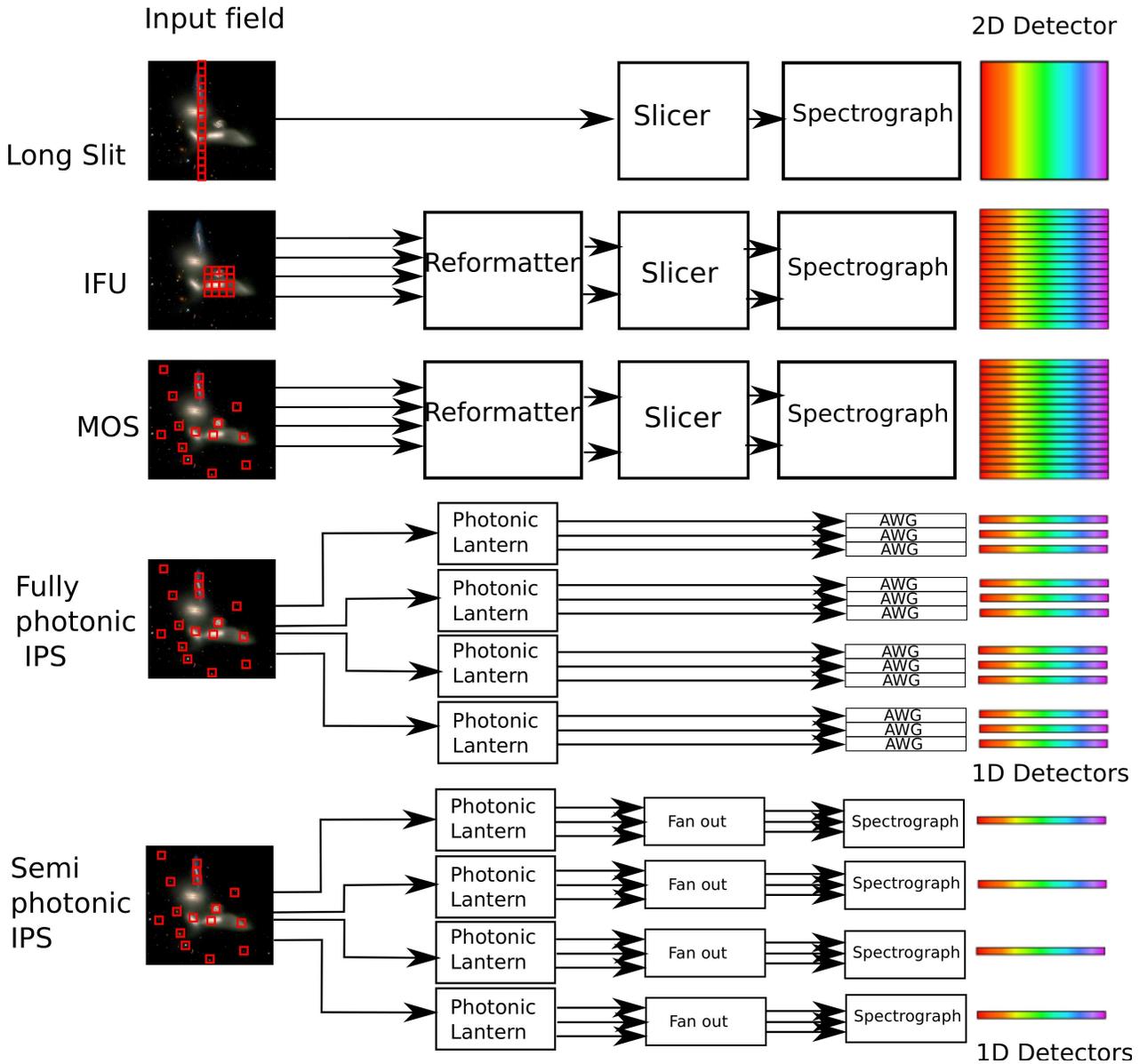}
		\caption{An illustration of conventional slicing and Photonic slicing. All methods sample an area of the same size (e.g. the same number of spaxels). The three conventional methods reformat the input and disperse it, producing one spectrum per spaxel. The fully-photonic option takes the each input spaxel and splits it into individual modes using a photonic lantern. Each of these modes is fed into an AWG to produce a spectrum. These then need to be recombined and summed to produce the spectrum for the spaxel. The semi-photonic option uses the same photonic lantern, but this is then reformatted into a long slit and fed into conventional spectrographs.}
	\label{fig:alt_slicing}
	\end{center}
\end{figure*}

A long slit can be thought of as a series of spaxels joined to form a rectangle of size 1$\times N$, where \textit{N} is the total number of spaxels. In IFS, the field is equivalent to  a series of slits (each composed of linked spaxels) joined so the total number of spaxels is $N=N_{x} N_{y}$, where $N_{x}$ is the number of spaxels in the $x$ direction and $N_{y}$ in the $y$ direction. MOS can be thought of as the same number of spaxels, $N$, distributed throughout the field of the telescope and brought together to form a long slit. See Fig. \ref{fig:alt_slicing} for an illustration of this.

\begin{figure}
		\includegraphics[width=84mm]{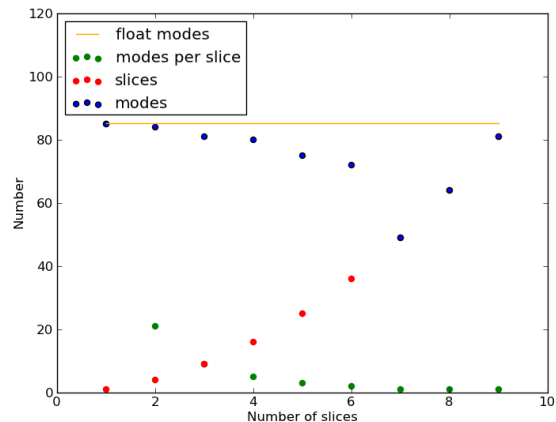}
		\caption{An example of the number of modes generated from a single spaxel on an 8m telescope of  0.5'' FWHM seeing at $\lambda$ = 1650nm. Fixing the size of the spaxel to the FWHM gives a single spaxel (here number of slices = 1), this spaxel is large and contains many modes. Slicing the spaxel produces smaller spaxel sizes, but larger numbers of them (the number of spaxels is the slices squared). This results in the same total number of modes in the area (the horizontal yellow line). The variation in the blue is due to the integerisation of modes within individual slices. Note that the three final red points lie under the blue ones.}
		\label{fig:awg_modes}
\end{figure}

An important consideration is the sampling of the field. From Fig. \ref{fig:awg_modes} and equation \ref{eqn:modes} it can be seen that the number of modes produced is dependent on the overall size of the field, not the individual spaxel size (as the number of modes per spaxel is proportional to the square of the spaxel size). This means the number of components for the IPS (and hence the approximate size of the instrument) required for the instrument will not depend on the sampling scale.  \newline
However, the amount of spatial information and throughput will depend on the 
sampling scale. Although it 
might appear best to reduce the spaxel size and the number of modes,
this will reduce the coupling efficiency and throughput\citep{Corbett2006}.  At the other extreme the use of very large fibres would result in loss of spatial information. A balance must be found between throughput and spatial resolution. We do not investigate this fully here as it does not affect the total number of modes in the field or the required number of detector pixels. Thus we choose to make our spaxel size equal to the FWHM of the seeing.

In order to calculate the scale length of the instrument we take the cube root of the volume of the instrument. We can calculate this from the number of spaxels in the total field passed to the spectrograph
\begin{equation}
	\mbox{scale length} = \sqrt{N M(\lambda_{min})^{P-1} L_{x} L_{y} L_{z}}. 
\end{equation}

Where \textit{M} is a function of the shortest wavelength in the spectrograph ($\lambda_{min}$), in order to account for all spatial modes. $L_{x}$, $L_{y}$ and $L_{z}$ are the lengths of an individual component spectrographs or AWGs in the $x$, $y$ and $z$ direction respectively which will be defined in the next two sections. \textit{P}=1 and \textit{P}=2  represent the semi-photonic and photonic cases respectively.

 \section{The semi-photonic IPS (P=1)}
 \label{sec:echelle}
 
The semi-photonic case involves taking an individual spaxel and using a re-arranged PL to form a diffraction-limited slit, which is then dispersed by bulk optics \citep{Bland-Hawthorn2010}.  

\subsection{Model Geometry}
\label{ech:geometry}

Slicing the input of a spectrograph has already been examined theoretically in \citep{Allington-Smith2009a}. The paper took existing instruments and sliced the input, either adding the slices to the length of the slit, or placing them into replica spectrographs. It showed that slicing could  result in an instrument with a slightly smaller overall volume, though the instruments sliced to the diffraction-limit were shown to be larger than their counterparts due to the extra components required. This is important to us as we showed that photonic spectroscopy is similar to image slicing to the diffraction limit in section \ref{sec:modes}. \newline
As conventional image slicing has already been examined we restrict ourselves to examining only the IPS concept, which takes each individual spaxel (not a number of them) and separates it into a single spectrograph. \newline
We will be using the modified model from \citep{Allington-Smith2009a} described in \citep{Harris2012} and adding this to our results. As the input to each spectrograph now depends on the number of modes per spaxel we shall be setting the length of the slit to the number of modes (equation \ref{eqn:modes}) instead of $n_{y}$ in the previous papers. 

\subsection{Semi-photonic model limits and Calibration}
\label{sec:ech_cal}

In order to calibrate the model we use the same method as \citep{Allington-Smith2009a}, with the  \textit{S}=1 case oversizing using a multiplicative factor and the \textit{S}=2 case oversizing the spectrograph input beam. The scaling factors for our instruments can be found in Table \ref{tab:ech_scaling}.

 \begin{table}
 \centering
 \begin{minipage}{86mm}
  \caption{The scaling parameters for the semi-photonic versions of the conventional instruments. The scaling scenarios are described fully in \citep{Allington-Smith2009a} and \citep{Harris2012}.}
	\begin{tabular}{ | c | c | c | c | c | }  \hline 
		Instrument  &  \multicolumn{2}{|c|}{S =1}  &  \multicolumn{2}{|c|}{S =2} \\ 
		 & a (m) & b & a (m) & b \\ \hline
		GNIRS & 0.1 & 2.1 & 0.46 & 1.1   \\ 	
		CRIRES & 0.1  & 2.2 & 0.61  & 1.1\\  
		NIFS (J) & 0.1 & 7.0 & 0.86 & 1.1\\  
		SINFONI (H) & 0.1 & 7.0 & 0.86 & 1.1     \\  \hline 
	  
		IRMS & 0.1 & 8.0 & 2.06  & 1.1  \\ 
		IRIS & 0.1 & 10.0 & 1.46 & 1.1 \\ 
	\end{tabular} 
	\label{tab:ech_scaling}
\end{minipage}
\end{table}

\section{The fully-photonic IPS (P=2)}

For our fully-photonic model we shall concentrate on modelling the size of the AWGs, not the components that feed them. We shall include a factor for our detector sizes.

\label{sec:awg}
\subsection{Model geometry}

For the fully-photonic model we first need to consider  the geometry of AWGs. These are available in many different variations, especially with respect to the geometry which generates the path difference between waveguides (e.g. S-bend, circular, horse-shoe). To keep the toy model simple we have chosen a reflective AWG  \citep{Peralta2003,DePeralta2004} using a Rowlands Circle arrangement for the Free Propagation Region (FPR). We have removed the bend at the end of the waveguide array for simplicity. Because of this it looks almost identical to a conventional double-pass Echelle spectrograph.

\begin{figure*}
	\begin{center}
		\includegraphics[width = 168mm]{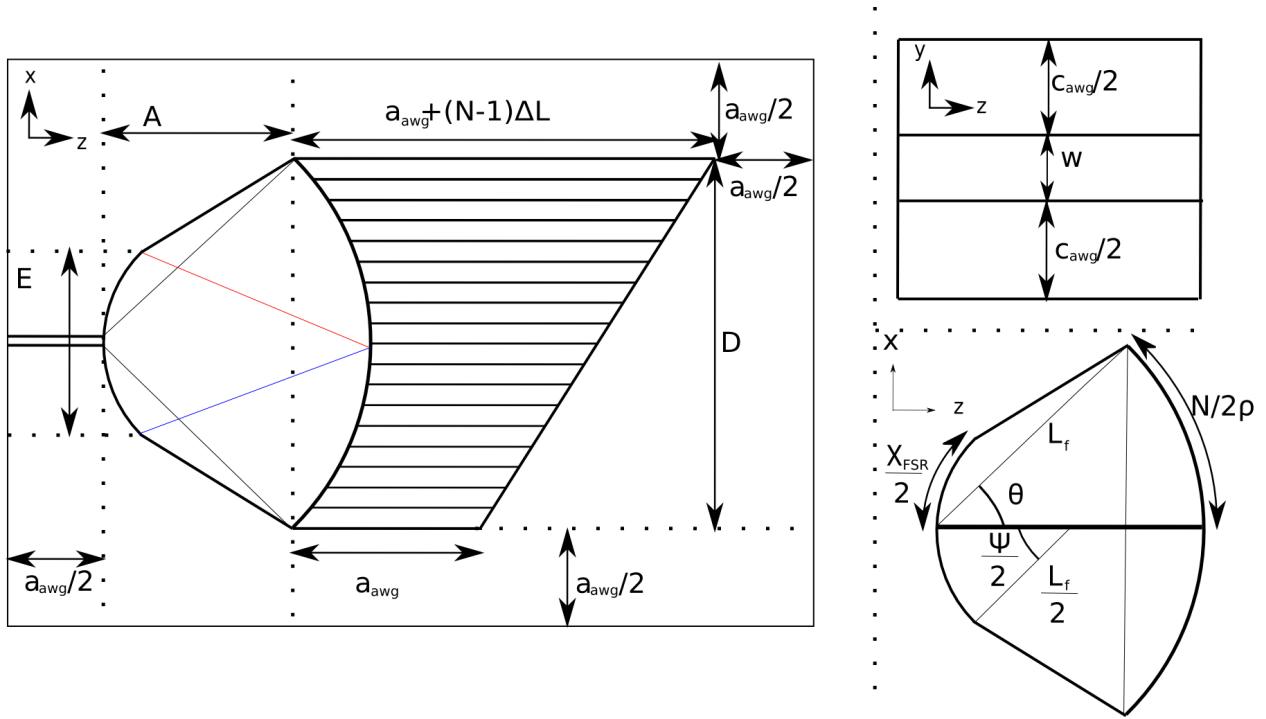}	
		\caption{The fully-photonic model. The left image shows the x-y view of the AWG, with the (left to right) input fibre, Free Propagation region and waveguides. The top right image shows the side view of the AWG model, with the top and bottom cladding layers and the layer containing the waveguides in the centre. The bottom right image is an enlargement of the FPR which is a Rowlands Circle arrangement.}
		\label{awg_model}
	\end{center}
\end{figure*} 

Using the definitions in Fig. \ref{awg_model} we arrive at the following equations for the size of the AWG model

\begin{eqnarray}
	L_{x} &=& \left( \rm max(D,E) + a_{awg}   \right)b_{awg}  \label{eqn:linear_x} \\
	L_{y} &=&  (c_{awg} +  w)b_{awg}  \label{eqn:linear_y} \\
	L_{z} &= &  \left( a_{awg} + A + (N_{wg}-1)\Delta L  \right)b_{awg}. \label{eqn:linear_z} 
\end{eqnarray}

Here \textit{A} is the $x$-length of the FPR,  $\Delta L$ the length difference between adjacent waveguides to achieve the required order for a given central wavelength ($\lambda{c}$),\textit{D} is the length containing the waveguides (analogous to the illuminated length of the grating in a standard echelle grating), $E$ the \textit{x} length of the detecting surface, \textit{w} is the waveguide diameter. The oversizing parameters  $a_{awg}$, $b_{awg}$ and $c_{awg}$ parameterise the extra size required to implement a practical device. 

First we calculate the appropriate dispersion order, $m$, in terms of  the free spectral range  (FSR, $\Delta \lambda_{FSR}$) for an AWG 

\begin{equation}
	m =  \frac{\lambda_{min}}{\Delta \lambda_{FSR}}.\\
	\label{eqn:fsr}
\end{equation}

Setting $D = N_{wg} / \rho$, this can be combined with equation \ref{eqn:awg_res} where  $\rho$ is the density of waveguides (analoguous to the ruling density of a conventional disperser) to give

\begin{equation}
	D = \frac{C R}{m \rho}. 
\end{equation}
The physical extent of the FSR in an arrayed waveguide grating  is $X_{FSR} = (\lambda_{min} L_F  \rho/ n_{s} $), where $n_{s}$ is the refractive index of the slab and $L_F$ the length of the free space propagation region. Combining with geometrical arguments gives

\begin{equation}
	E = L_{F} \sin \left( \frac{\Psi}{2} \right) = L_{F } \sin \left( \frac{\lambda_{min} \rho}{n_{s}} \right).
\end{equation}

Where $\Psi$ is defined in Fig. \ref{awg_model}. In order to calculate $\Delta L$ we make use of the equation for calculating the central wavelength of the AWG 

\begin{equation}
	\Delta L = \frac{\lambda_{c} m}{n_{c}}
	\label{eqn:delta_l}
\end{equation}
where  $n_{c}$ is the refractive index of the waveguides, the central operating wavelength is
 $\lambda_{c}$ =$\lambda_{min} + \Delta \lambda_{FSR}/2$, so $A$ can be calculated from geometry as

\begin{equation}
	A = L_F  \cos( \theta) = L_F \cos \left( \frac{N_{wg}}{2\rho L_{F}} \right).
	\label{A_eqn}
\end{equation}

where $\theta$ is defined in the figure and $N_{wg}$ is calculated using equation \ref{eqn:awg_res}. 
In order to calculate $L_F$ we make use of the fact the imaging requires the number of detector pixels to be able to adequately sample at the resolution required (equivalent to sampling of the Echelle model in \citep{Allington-Smith2009a}. To do this we take the dispersion relation

\begin{equation}
\left(  \frac{ \delta \lambda} {\delta x} \right) \simeq 
\left(  \frac{d \lambda}{dx} \right) \
= \frac{n_{s}}{L_{f}m \rho}.
\end{equation}

and  combine it with  equation \ref{eqn:fsr}, setting $\delta  x =N_{0} d_{p}$,  where $N_{0}$ is the oversampling and $d_{p}$ is the size of the pixels. We also take the equation for the spectral resolution $\delta \lambda$ = $\lambda_{min} / R$.  Minimising to obtain the maximum $L_{F}$ we find 

\begin{equation}
	L_{F} \geq \frac{n_{s} N_{0} d_{p} R \Delta \lambda_{\rm FSR} }{\rho \lambda_{min}^{2}}.
	\label{eqn:Lf}
\end{equation}

Finally we can calculate the number of pixels we need for the required resolution

\begin{equation}
	N_{P} = \frac{L_{\rm FSR}}{N_0 d_{p}} = \frac{\lambda_{c} L_{F} \rho}{ n_{s} d_{p}}.
\end{equation}

\subsection{Fully photonic model Limitations and Calibration}
\label{sec:calibration}

Astronomical spectrographs are usually designed to operate with a large free spectral range (typically several hundred nm). This is a problem for the IPS because conventional telecoms AWGs are designed with low free spectral range in order to deal with the discrete narrow band input from the telecoms industry.  For astronomy, single AWGs need to be redesigned to work in lower spectral orders by reducing the path difference between adjacent waveguides. This requires more waveguides to maintain the maximum theoretical resolution and an increase in  $L_{F}$ to maintain a practical one (see eqn \ref{eqn:Lf}). This produces very large AWG dimensions which cannot be manufactured  due to chip manufacturing size constraints \citep{Lawrence2010}. \newline

\begin{figure}
	\begin{center}
	\includegraphics[width = 84mm]{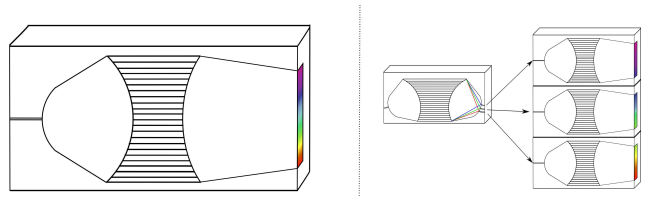}
		\caption{An illustration of the tandem and single AWG setup. The image on the left shows the conventional AWG dispersing the whole spectrum. The image on the right is the tandem configuration. The intial AWG (or other disperive optic) splits the light by wavelength (here to $\Delta \lambda_{FSR}/3$) and feeds the second set of AWGs. This has the advantage that each individual AWG can be smaller, though it requires more AWGs, additional components (feeding fibres) and is subject to extra loss of throughput. Note that the length of the output must be the same in both cases in order for the detector to sample adequately.}
		\label{fig:tandem_awg}
	\end{center}
\end{figure}

We wish to avoid this problem and retain a fully integrated design with no external optics.  As such we make use of the tandem AWG arrangement, where a primary AWG filters the light by wavelength into secondary AWGs, each encompassing a fraction of the original FSR (see Fig. \ref{fig:tandem_awg} and \citep{Takada2001}). This allows the individual component dimensions to be within manufacturing limits whilst allowing our full device to sample the correct FSR. It would also allow the AWG design process to remain similar to current specifications. 

Since no fullscale AWG instruments currently exist, we are required to use a bottoms-up approach to estimate its size. To do so we take an existing AWG acquired from Gemfire Livingstone. We use its known parameters and adjust the models produced dimensions until they match the real ones. In order to emulate \citep{Allington-Smith2009a}, we set the scaling parameters to two extremes. \newline

\begin{description}
	\item \textbf{S=1 : } Minimise $b_{awg}$=1.1,which yields a $a_{awg}$ = 10mm 
	\item \textbf{S=2 : }  Minimise $a_{awg}$ = 0mm, which yields $b_{awg} = 2.8$
\end{description}

For both scenarios we keep $c_{awg}$ = 0.7mm as the device is planar so the height should not change.

For simplicity we will not include the volume of the initial multimode fibre bundle, the photonic lantern or the housing of the instrument. We will however include a estimate for the size of the detector. This value is calculated assuming the size of a typical detector sub-system  including the cryostat. This is  estimated as   $10^{-7}m^{3}$ per detector pixel.

\section{Results}
\label{sec:results}

To estimate the uncertainties within our model, we use both of our oversizing options for both models and also vary $C$ between 1 (diffraction-limited) and 4 (the initial results obtained in \citep{Cvetojevic2009}) for our fully-photonic instrument. This gives us two extremes and allows for the current maximum of \textit{C} =1.6 achieved in  \citep{Cvetojevic2012}. 


%

To test our models we choose two sets of instruments. First those designed to represent current instrumentation on 8m telescopes. Then we test against instruments that have been designed for the TMT. We shall only investigate a single mode of operation for each instrument. This may be oversimplifying as most instruments are designed to operate at different resolutions and at different wavelengths by changing gratings or optics. It however keeps the model simple and IPS devices could be adapted to suit various purposes, this is discussed in the conclusions.   \newline

The first set of instruments are in current use on 8m telescopes and are intended to represent generic instrument types. We use the parameters in Table \ref{tab:parameters} to calculate the volume of the instruments. The instruments we have chosen are:

\begin{enumerate}

\item  Gemini Near InfraRed Spectrograph (GNIRS) on Gemini-North. The instrument has an overall wavelength range of 1.0-5.4$\mu m$, resolutions of between 1,700 and 18,000 and slit widths of between 0.1 and 1.0 arcseconds. It has an imaging mode, a long slit mode and originally an IFS mode (destroyed during maintenance at the telescope).
We will be comparing our photonic instrument to the long slit configuration. \newline

\item CRyogenic high-resolution InfraRed Echelle Spectrograph (CRIRES) is a high resolution spectrograph on the Very Large Telescope (VLT). It is designed to operate between 1.0-5$\mu m$, with a resolution of up to $10^{5}$. We have chosen it to illustrate a long slit high resolution spectrograph. \newline

\item Near-Infrared Integral Field Spectrometer (NIFS) on Gemini-North is our first IFU instrument. It is  designed to work with the Adaptive Optics system, over 0.9-2.4 $\mu m$. \newline
\noindent 

\item Spectrograph for INtegral Field Observations in the Near Infrared (SINFONI) on  the VLT is our second IFU instrument. It operates in the 1.1 to 2.45$\mu m$ range again with AO. \newline

\end{enumerate}


We have also chosen two hybrid instruments proposed for the Thirty Meter Telescope \citep{2010SPIE.7735E..70S}.  IRMS will employ 10 or more IFUs. Each one will have a 2 x 2 arcsecond squared field of view with a 50\% of the energy enclosed by 50 mas at wavelength 1$\mu$m, resulting in 1600 spaxels per IFU. IRIS has 3 IFU units, two of which will be lenslet arrays (for observing smaller fields) and one will be an image slicer (larger fields). Here we model the slicer, which has 88 mirror facets, but will keep the best resolution possible with our AWG model. As with the 8m instruments , the instrument scale lengths are fitted to values taken from the literature, see Table \ref{tab:parameters}. \newline

\begin{table*}
 \centering
 \begin{minipage}{140mm}
  \caption{Table of input parameters all instruments. Symbols except for $N_{IFU}$, the number of IFUs in the instrument, are explained in the text. The numbers are taken (and approximated from) \citep{Allington-Smith2009a} for GNIRS, \citep{Kaeufl2004}  for CRIRES, \citep{McGregor2003} for NIFS, \citep{Eisenhauer2003} for SINFONI, \citep{Eikenberry2006} for IRMS and \citep{LarkinTMT} for IRIS.}
	\begin{tabular}{ | c | c | c | c | c | c | c | c| c | }  \hline 
		Instrument  & $\chi$ & $N_{y} (N_{x})[ N_{IFU]}$ & Total Spaxels  & R & $\lambda_{c}$  & $\Delta_{FSR}$  & $\rho$  &    Vol  \\ 
 		  & (") &  &  &   & (nm)  &  (nm) & ($mm^{-1}$) & $(m^{3})$     \\ \hline
		GNIRS & 0.3 & 330(1) & 330 & 5,900 & 1,650 & 400 & 31.7 & 2.00  \\ 	
		CRIRES & 0.3 & 200(1) & 200 & 100,000 & 1,650 & 48 & 31.6 & 3.00  \\  
		NIFS (J) & 0.1 & 30(29) & 870 & 6,050 & 1,250 & 600 & 600 & 2.75 \\  
		SINFONI (H) & 0.2 & 32(32) & 1,024 & 3,000 & 1,650 & 400 & 128.57 & 2.75   \\  \hline 
	  
		IRMS & 0.1 & 40 (40)[10] & 16,000 & 10,000 & 1,200 & 400 & 128.57 & 16.00  \\ 
		IRIS & 0.1 & 60(60)[1] & 4,000 & 8,000 & 1,200 & 400 & 310 & 55.00 \\ 
	\end{tabular} 
	\label{tab:parameters}
\end{minipage}
\end{table*}

\begin{table*}
 \centering
 \begin{minipage}{140mm}
  \caption{Table of the resulting scale lengths of the respective fully-photonic instruments. The total number of modes in the whole instrument is shown first, with the next four columns showing the respective sizes for the model with no detector. This is followed by the model with detector. All of the scale lengths are normalised to the cube root of the volume in Table \ref{tab:parameters}. The AWG model uses a waveguide separation of $\rho$ = 200$mm^{-1}$.}
	\begin{tabular}{ | c | c | c | c | c | c | c | c | c | c | | }  \hline 
		\multicolumn{2}{|c|}{} & \multicolumn{8}{|c|}{Normalised Scale length} \\
		 & Total    & \multicolumn{4}{|c|}{No detector } & \multicolumn{4}{|c|}{ With detector }   \\ 
	  Instrument& number   & \multicolumn{2}{|c|} {\textit{C}=1}   &  \multicolumn{2}{|c|} {\textit{C}=4} & \multicolumn{2}{|c|} {\textit{C}=1}   &  \multicolumn{2}{|c|} {\textit{C}=4} \\ 
		&  of AWGs  & \textit{S}=1 & \textit{S}=2 & \textit{S}=1 & \textit{S}=2 & \textit{S}=1 & \textit{S}=2 & \textit{S}=1 & \textit{S}=2  \\ \hline
GNIRS & 13,000  & 0.49 & 0.87 & 1.46 & 1.55  & 0.51 & 0.89 & 1.47 & 1.55  \\ 
CRIRES &  6,300  & 0.50 & 0.90 & 1.19 & 1.32  &0.60 & 1.09 & 1.22 & 1.42  \\ 
NIFS & 8,900  & 0.79 & 1.45 & 1.66 & 1.93  & 0.80 & 1.46 & 1.66 & 1.93   \\ 
SINFONI & 28,000  & 0.39 & 0.64 & 1.36 & 1.39  & 0.39 & 0.66 & 1.36 & 1.39   \\ \hline

IRMOS & 520,000 & 1.69 & 3.08 & 3.53 & 4.09  & 1.71 & 3.11 & 3.53 & 4.11   \\ 
IRIS & 110,000 & 0.72 & 1.07 & 1.34 & 1.48  & 0.73 & 1.08 & 1.35 & 1.49   \\ 

	\end{tabular} 	
	\label{tab:results_vol}
\end{minipage}
\end{table*}

\begin{table*}
 \centering
 \begin{minipage}{140mm}
  \caption{Further information on the fully-photonic model. The AWG model uses a waveguide separation of $\rho$ = 200$mm^{-1}$. }
	\begin{tabular}{ | c | c | c  | c | c |}  \hline 
	Instrument & Modes  & Total detector &  Reference Instrument \\
	 &  per spaxel &  pixels /(10$^{6}$) & pixels /(10$^{6}$) \\ \hline
		GNIRS & 39 &  60.69   & 1.05 \\ 
		CRIRES &  31  & 47.31   & 2.10 \\ 
		NIFS & 10  &  112.43  & 4.19 \\ 
		SINFONI & 27  &  66.50  & 4.19 \\ \hline

		IRMOS & 32  & 6263.44   & 83.89 \\ 
		IRIS & 32  & 1127.42   & 16.78 \\ 

	\end{tabular} 	
	\label{tab:results_det}
\end{minipage}
\end{table*}

\begin{table*}
 \centering
 \begin{minipage}{140mm}
  \caption{Results for the semi-photonic model. All of the scale lengths are normalised to the cube root of the volume in Table \ref{tab:parameters}. The semi-photonic model uses $\rho$ stated in Fig. \ref{tab:parameters}. * The number of detector pixels assumes that the modes in each spaxel can be reduced onto the detector appropriately, which may not be the case.}
	\begin{tabular}{ | c | c | c | c | c |}  \hline 
		 & Number & \multicolumn{2}{|c|}{Normalised Scale }   & Number of \\
		 Instrument & of replica & \multicolumn{2}{|c|}{Length }  & detector \\  
		 & Spectrographs  &S = 1 & S = 2 & pixels* /(10$^{6}$) \\ \hline
		GNIRS & 330 & 4.30 & 5.30 &1.05  \\ 
		CRIRES & 200 & 3.06 & 4.66  & 2.10 \\ 
		NIFS & 870 & 9.55 & 9.54 & 4.19 \\ 
		SINFONI & 1,024 & 8.62 & 9.36 & 4.19 \\ \hline

		IRMOS & 16,000 & 26.07 & 32.48 & 83.89  \\ 
		IRIS & 3.600 & 8.49 & 9.20 & 16.78 \\ 

	\end{tabular} 	
	\label{tab:results_ech}
\end{minipage}
\end{table*}

Table \ref{tab:results_vol} shows the resulting parameters in the fully-photonic case. The total number of AWGs required are shown in the second column, this number will be in the tens of thousands for 8m instruments and the hundreds of thousands for the 30m instruments. The large number of AWGs requires rigorous quality control to test the large number of individual components. This may be of advantage though, as the individual AWGs should be less prone to flexure and, due to their modular nature, are better suited to mass production and upgrades and expansion to suit cashflow. Note that the size predictions do not include provision for mounting hardware required to support the instrument components or to provide a suitable controlled environment. \newline
The next eight columns show the different resulting normalised scale lengths of the instrument. Note that this is the scale length of the overall instrument, not the individual components. 
The first four are the scale lengths without provision for the detector size and show that the total size of the 8m instruments will be on the same order as the conventional instrument. If the diffraction limit can be achieved the resulting instruments are smaller for all scenarios, with the exception being NIFS with the \textit{S}=2 scaling. If the diffraction-limit cannot be achieved the instruments will have a scale length larger than the original instruments. 
The results for the 30m instruments are similar to the 8m ones, with the \textit{S}=1 scaled case of IRIS being slightly smaller and the rest being slightly larger. \newline
The second four include the provision for detector and show similar results, though the scale lengths are increased slightly as expected. This shows though the size of the additional detector pixels (discussed later) will not pose a significant size restriction on the instrument. \newline

Table \ref{tab:results_det} shows the results corresponding to the number of modes per spaxel and hence the requirements in terms of detector pixels. The second column shows that all of the instruments will have around 30 modes per spaxel, with the exception of NIFS, which will have 10. This causes problems with oversampling in the fully-photonic model due to each mode needing to be sampled using two detector pixels per resolution element (Nyquist sampling). There needs to be some way of combining the individual spectra to stop massive oversampling (shown in column 3). This will be discussed later, but will probably involve additional components, increasing the size of the instrument. 

Table \ref{tab:results_ech} shows the results from the semi-photonic model. The second column shows the number of replica photonic instruments will be 10$^{3}$-10$^{4}$. As such using the semi-photonic method will require mass production of the replica spectrographs which is not common in the astronomical instrumentation community. The alternative is to put many spaxels in the same spectrograph and would require a balance between redundancy and overlarge components for this version is to work. \newline 
The next two columns show that using the semi-photonic model for the instrument will result in much larger instruments. This  matches with the results of \citep{Allington-Smith2009a}, where as the input was sliced more the instrument tended to get bigger. As stated in section \ref{sec:echelle} we are slicing the instrument to the diffraction-limit in the spatial direction and then separating each spaxel into a separate spectrograph, which imposes huge redundancies. The number of required detector pixels are shown in the final column  for this we are assuming that all the detector separate modes can be combined onto a linear detector, which may not be possible.

 %
 


 \section{Modifications to  Integrated Photonic Spectrographs}
\label{sec:cross_disp}

From the results already discussed, it is clear that  IPSs in their current state offer little or no advantage in terms of size and detector pixels when compared with existing instruments on large telescopes or those planned for ELTs. However it is possible to envisage modifications to the fully-photonic device which would make it possible to exploit the unique 
features of photonic spectrographs. One such scheme is already being studied \citep{Cvetojevic2012}.

\subsection{Multiple-input Arrayed Waveguide Gratings}

\begin{figure}
	\begin{center}
		\includegraphics[width = 84mm]{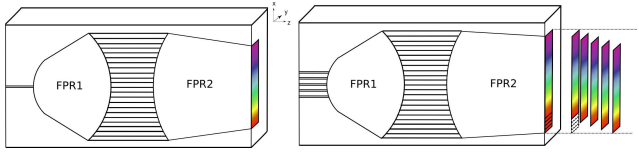}
		\caption{Cutaway diagram illustrating the difference between the single and multiple input versions of the AWG. To the left is the single input, which would make use of a linear detector array to sample the output spectrum. To the right the multiple input version. Here five inputs produce five separate spectra that overlap at the output of the AWG. This has a couple of implications, first the spectra would need to be cross dispersed in order to be sampled and secondly the end of the second FPR would need to be larger (though not the input which would remain the size of the waveguides). The path difference in the waveguides is not illustrated in this diagram for simplicity. }
		\label{fig:multi_linear}
	\end{center}
\end{figure}

\begin{figure*}
	\begin{center}
		\includegraphics[width = 168mm]{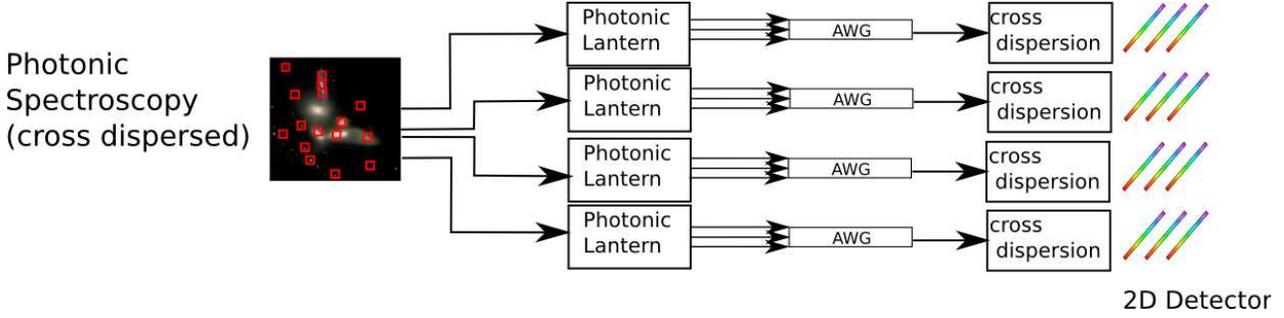}
		\caption{The multi-input model for the cross dispersed system. Each spaxel from the input field is fed into a photonic lantern. The output SMFs are fed into AWGs, with multiple fibres in each AWG. The output from these AWGs is then cross dispersed onto a 2d detector.}
		\label{fig:cross_diag}
	\end{center}
\end{figure*}

So far we have restricted ourselves to one input per AWG (i.e. one mode per AWG), which as shown in previous sections requires many AWGs. Placing multiple inputs per AWG would reduce the dependence on equation \ref{eqn:modes}, at the extreme eliminating it entirely if all the inputs could fit into a single AWG.  In order to introduce these extra inputs additional fibres are placed at different positions on the input FPR. This introduces a path difference between each input with respect to the central one (see Fig. \ref{fig:multi_linear} and \ref{fig:cross_diag}). This path difference is carried through the system and results in the output spectrum of each mode being shifted in the dispersion direction at the output. To remove the overlap between spectra, it will be necessary to introduce  cross dispersion.

The AWG also produces multiple diffraction orders (as with a conventional grating) so we also need to make sure the inputs all lie within a region half that of the  FSR of the central input. This is to stop the same wavelength from different orders  lying in the same position in the linear output, resulting in the cross dispersed spectra lying the same position on the detector. The refractive index change in fused Silica is not great enough to disperse the light by polishing the AWG at an angle. In order to separate the spectra, the outputs needs to be cross-dispersed using conventional optics  \citep{Cvetojevic2009} . This means that each AWG will need a two-dimensional detector, a dispersive element and collimating and camera lenses. Here we look at the relative advantages of using cross dispersion in the system.

\subsection{Adding cross-dispersion}

To cross-disperse, we need additional optics, which means that the device is no longer fully integrated, potentially making manufacture and maintenance more difficult, but reducing the number of AWGs required for the device.
We construct a new toy model to see how the scale length of a cross dispersed system (multiple inputs) compares to one with linear arrays (single inputs). For this section we have modified our fully-photonic model so it is no longer reflective and the output of the second FPR outputs is linear (e.g. Fig. \ref{fig:multi_linear} and \citep{Lu2003}). The first FPR is still in its original shape so as to allow  the multiple inputs. Changing the model like this will affect the overall size of the instrument (due to the difference in AWG design), but will still allow us to examine the relative sizes of the two scenarios. \newline

We retain the single input model for our comparison and use the length values calculated in previous sections. As such the scale length of the instrument (with all modes) is still $N_{awg} L_{x}L_{y}L_{z}$, where $N_{awg}$ is the number of AWGs. 

\begin{figure*}
	\begin{center}
		\includegraphics[width = 168mm]{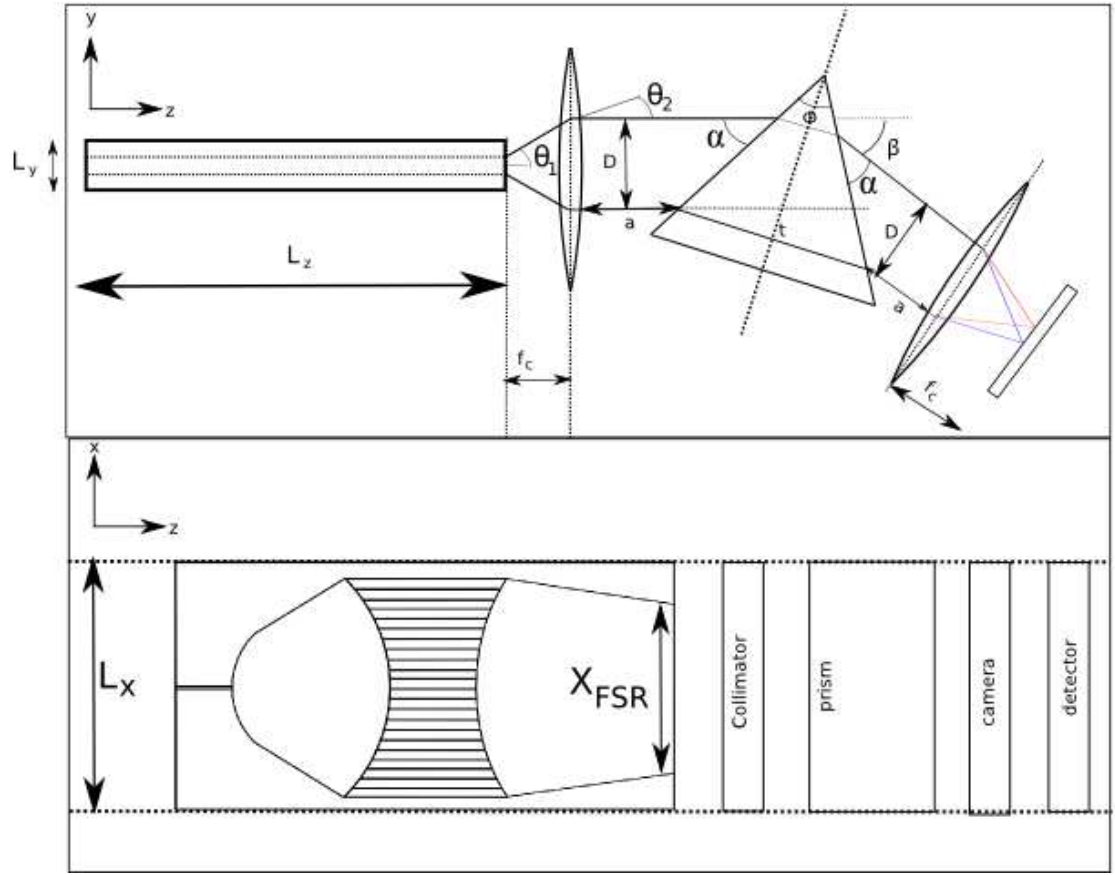}
		\caption{The new model for the AWG. This allows multiple inputs to the AWG and includes cross dispersion at the output in order to separate overlapping spectra.}
		\label{fig:xdisp}
	\end{center}
\end{figure*}

The calculation of the size of the multiple-input AWG option follows that for the 
The cross dispersion option requires the AWG, a collimator, prism and then camera in front of the detector (see Fig. \ref{fig:xdisp}). The equation for the volume of the system now becomes $ N_{awg} L_{x}L_{y}L_{z}$, where $N_{awg}$ = $N/N_{i}$, the total number of spaxels divided by the number of inputs per AWG and the dimensions being defined below.


We start by examining the output end of the AWG. For a single input the \textit{x} length of the AWG system would be the same as described in equation \ref{eqn:linear_x},with \textit{E} will now be $X_{FSR}$ as we have flattened out the output. Adding extra inputs such as the ones illustrated in Fig. \ref{fig:xdisp} will increase this \textit{x} length. The maximum distance between inputs must be less than $X_{FSR}$, to avoid the same wavelength in a different orders lying on the same position. For simplicity we assume evenly spaced inputs, which when combined with the previous condition yields equation \ref{eqn:lx_xdisp}. We set the maximum number of inputs to be $X_{FSR}$ / $D_{input}$, where $D_{input}$ is the diameter of the input fibre (here set to 125$\mu$m). \newline
To calculate the \textit{y} length, we must consider how the spectra are to be cross dispersed. We need to make sure that the output beam from the system is collimated. To do this we make sure the output angle of the collimator is smal: $\theta_{2}$ =  \textit{w}/ $2 f_{c}$ $<$  0.01\degr. where $\theta_{2}$ is the divergence in the collimator and $f_{c}$ is the focal length of our collimator. As our system is diffraction-limited, the diameter of our collimated beam is

\begin{equation}
	D = \theta_{1} f_{c} \approx \left( \frac{\lambda_{max}}{w} \right) f_{c}
\end{equation}

For  cross-dispersion we use a prism, athough a grating could also be used. We need to work out the required resolution of the prism, which is proportional to the number of inputs (e.g. as the number of inputs increases the FSR decreases by that factor). This gives

\begin{equation}
	R_{x} = \frac{N_{i} \lambda_{min}}{\Delta \lambda_{FSR}}.
\end{equation}
where $R_{x}$ is the resolution of the cross dispersed system. We can then combine this with the equation for the resolving power of a diffraction-limited prism \citep{Foy2005} to yield

\begin{equation}
	t >  R_{x}   \left( \frac{d \lambda}{dn} \right) = \left(\frac{N_{i} \lambda_{max}}{\Delta \lambda_{FSR}}  \right)  \left( \frac{d \lambda}{dn} \right) .
	\label{eqn:t}
\end{equation}
where $t$ is the path difference between the upper and lower rays in the prism. In order to account for all wavelengths we must use the maximum value of the material dispersion, $d \lambda / dn$ within the required wavelength range). The vertex angle of the prism is

\begin{equation}
	\alpha = \arcsin \left( \frac{2 D}{nt} \right).
\end{equation}
where $\alpha$ is the angle the prism makes to the collimated beam and n is the refractive index of the prism. This allows the calculation of the vertex angle of the prism

\begin{equation}
	\phi = 2 \arcsin \left( \frac{t}{2D} \sin \alpha \right).
\end{equation}
The output angle of the prism
\begin{equation}
	\beta = \pi - \phi - 2 \alpha.
\end{equation}
We can then calculate $L_{y}$ from Fig. \ref{fig:xdisp}, giving equation \ref{eqn:ly_xdisp}. Finally from above and from the geometry in Fig. \ref{fig:xdisp} we have

\begin{eqnarray}
	L_{x} &=& MAX \left( X_{FSR} \left( \frac{2 N_{i} -1}{N_{i}}  \right),  D \right) + 2a_{awg} \label{eqn:lx_xdisp} \\
	L_{y} &=& \frac{D}{2} + t \sin \delta + a \sin \beta + MAX \left( f_{c} \sin \beta, \frac{D}{2} \right) \label{eqn:ly_xdisp} \\
	L_{z} &=& L_{z} + f_{col} + a + t \cos \delta + \left( a + f_{c} \right)  \cos \beta. \label{eqn:lz_xdisp} 
\end{eqnarray}

\subsection{Results for cross-dispersed multiple-input AWG}

We now run the simulation for all the instruments detailed in section \ref{sec:results} using the model above and fused silica as the glass in our prism. We also set the maximum value  of \textit{t} to  30cm, to represent sensible limits for the prism size.

\begin{figure*}
	\begin{center}
		\begin{tabular}{cc}
			\includegraphics[width=84mm]{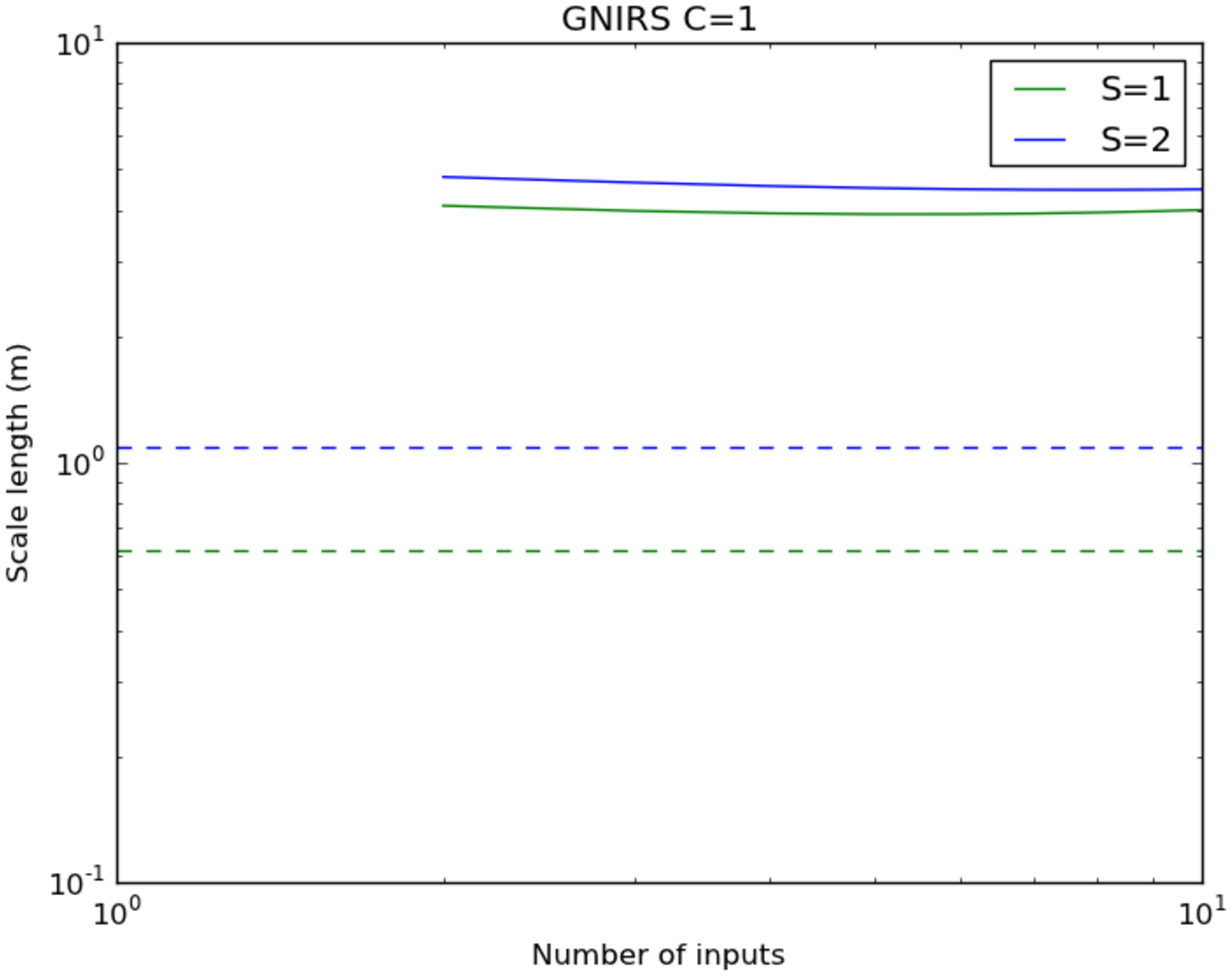}&
    			\includegraphics[width=84mm]{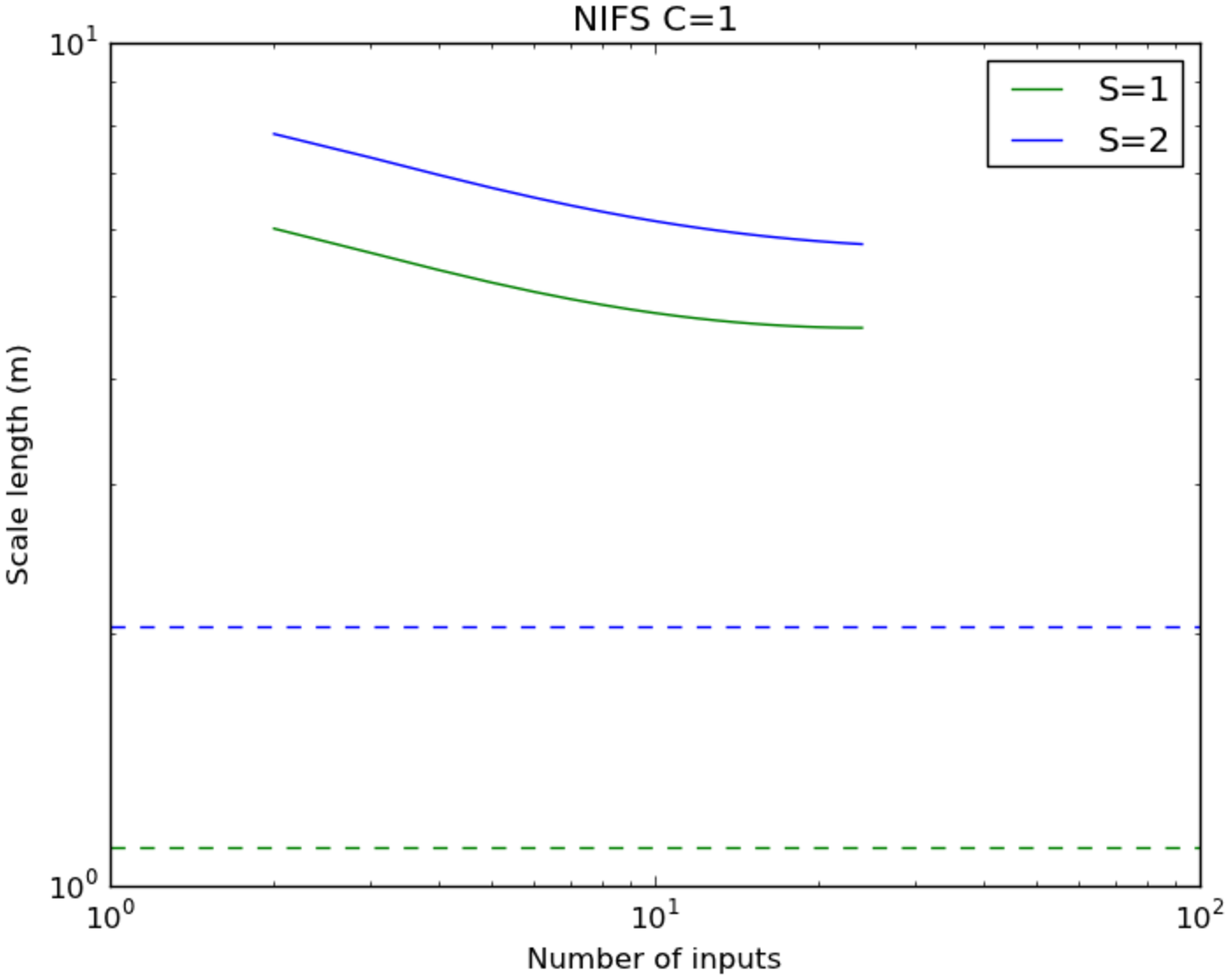}\\
			\includegraphics[width=84mm]{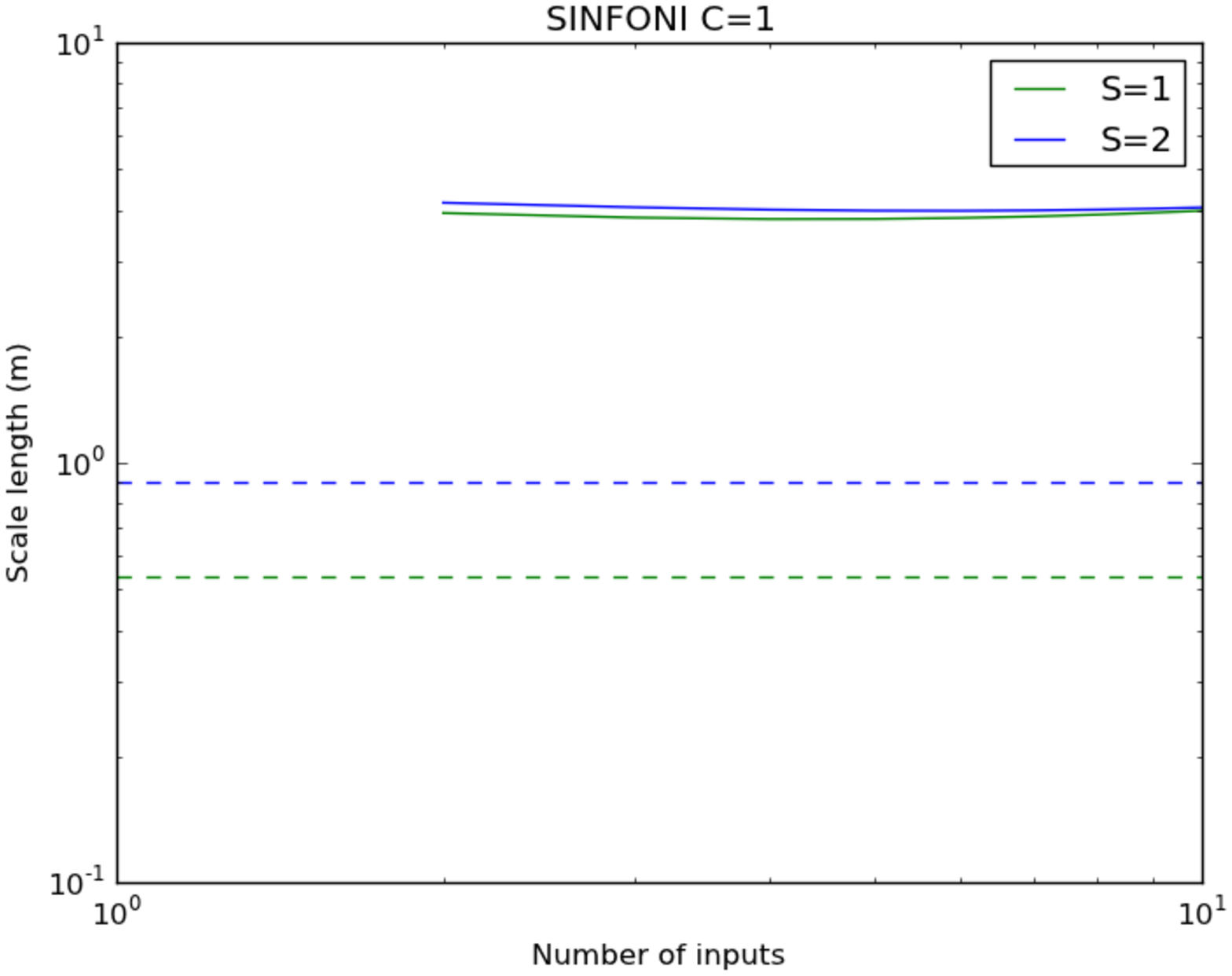}&
  			\includegraphics[width=84mm]{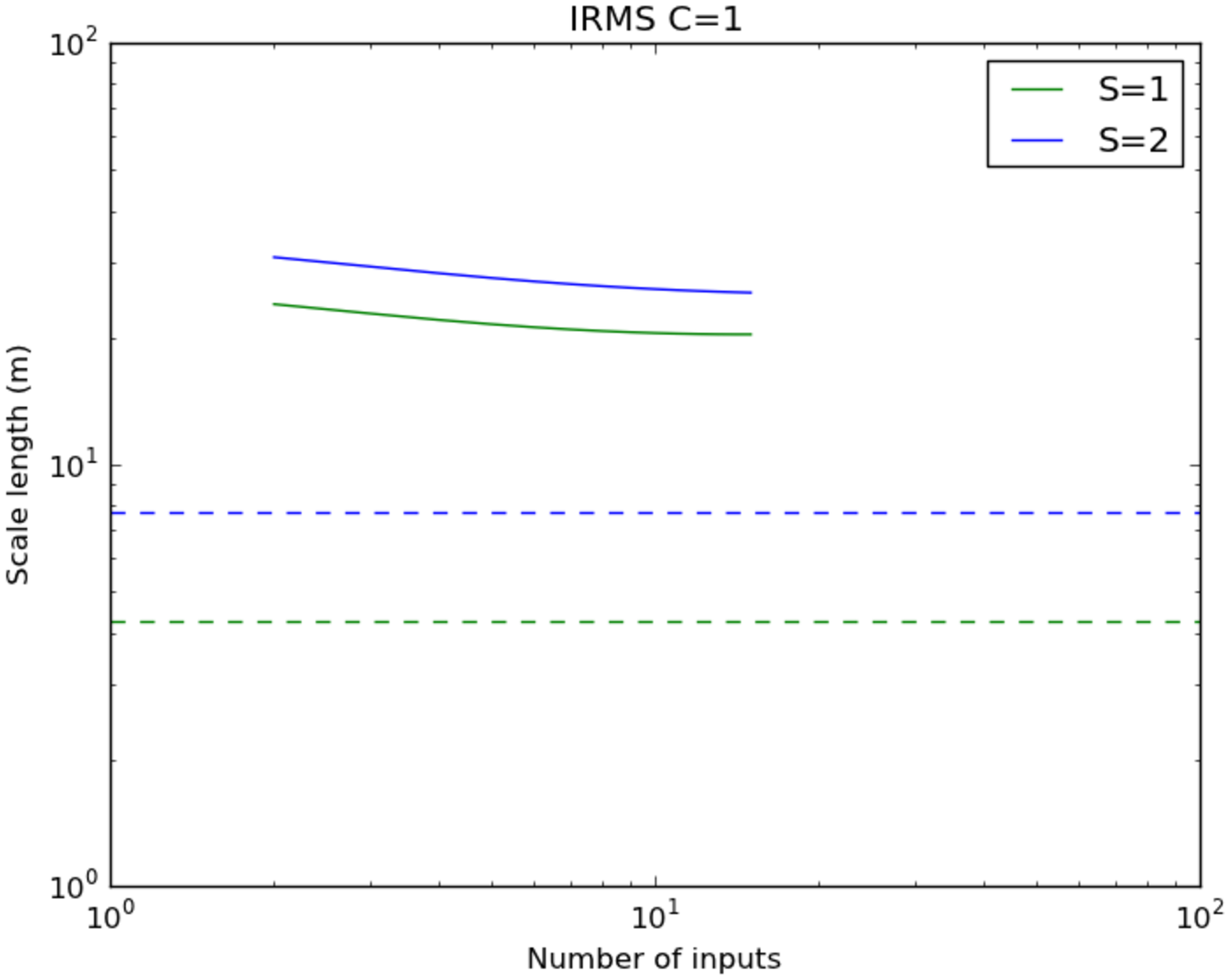}\\
			\includegraphics[width=84mm]{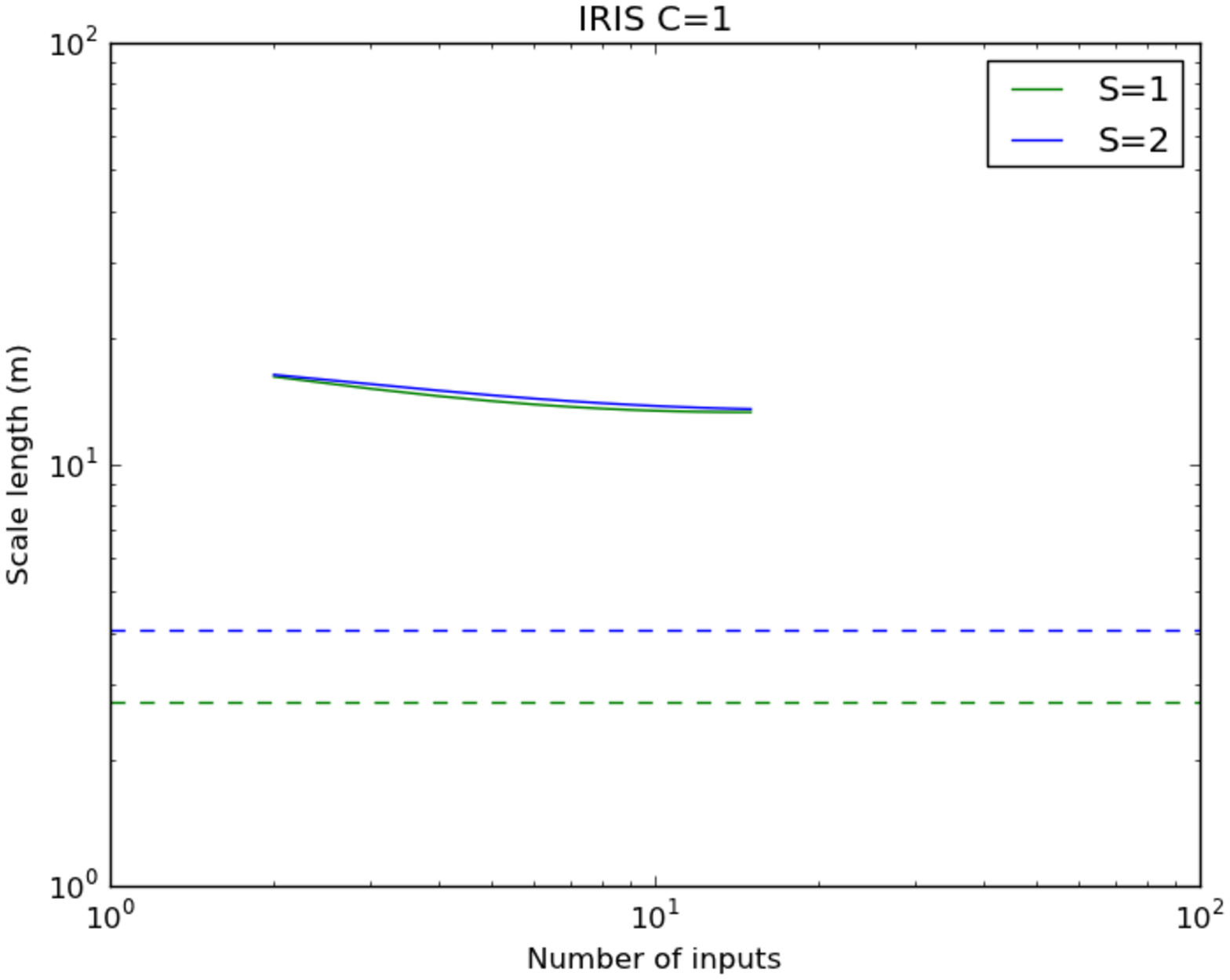}&
    			\\
  		\end{tabular}
  		\label{fig:cross_results}
		\caption{The resulting scale length due to varying the number of inputs to each AWG on each instrument. The different scaling cases are show in green and blue with the dashed horizontal line indicating the scale length of the single input instrument. From the figure you can see that all the results will produce larger instruments than the single input case. The result for CRIRES is omitted as no sensibly sized prism could be found with sufficient resolution to cross disperse the outputs.}
	\end{center}
\end{figure*} 

By imposing our limit on \textit{t} we can see in  Fig. \ref{fig:cross_results} that the number of inputs per chip is limited to the tens for all the resulting graphs due to equation \ref{eqn:t}. The potential advantage of this is all the resulting modes from a single spaxel could be fed into one AWG, meaning each one could be isolated. \newline
We can also see that though the instrument size decreases (particularly for NIFS, IRMS and IRIS) as more inputs are added all instrument sizes will be much larger than the single input version. Existing results have only put around 10 inputs on a chip and then cross dispersed by the IRIS2 instrument \citep{Cvetojevic2012}, which fits with out results. There is no result for CRIRES as the prism would have to be too large to have sufficient resolution.
 
 Not shown in the resulting graphs are the numbers of pixels required for the instruments, which would be of the same order or greater for this new setup.
 
It should be noted that we have used a prism in our example, which is usually used for lower resolution cross dispersion. The alternative is to use a grating, though this would  work in a similar way. Taking the equation for FSR and combining it with a diffraction-limited grating ($R_{x} = m \rho W$) yields

\begin{equation}
	N_{i} = \rho W.
\end{equation}
Showing the grating size (related to $W$) will increase as the number of inputs increases (given a maximum ruling density).


\subsection{Other instrument options}

In its present form it is clear that trying to compete with large IFU style instruments is not a viable option. As shown in \citep{Harris2012} the areas providing the greatest advantage would be small or diffraction-limited telescopes, preferably operating at longer wavelengths with instruments that only require a small field of view. There is potential for applications in solar system science, planetary and stellar science and studies of individual stars in galaxy populations

Another option is to use multiple single-input AWGs, but to combine the output onto a single linear detector array to reduce the number of detector pixels by a factor equal to the number of modes per spaxel. This would greatly reduce the cost of the detector system and bring the benefit of adaptability of a one-dimensional  detector array to the output focal surface of the AWG. This would only be possible if the pixels had a large aspect ratio. This might incur a penalty in terms of extra detector noise; and the number of AWGs is not reduced. This option is currently under investigation. 

A further option is to reduce the number of modes that are extracted from the input multimode light to produce an acceptable tradeoff between cost and performance defined as a combination of  throughput, spectral resolution and field size. Options include..

(a)  Restrict the number of single-mode fibres output from the photonic lantern with a consequent loss of throughput. This may  be acceptable  because the population of excited modes is not likely to be uniform \citep{Corbett2006} but will reduce as a function of mode number to a  cutoff value at high order. Thus the overall performance of the system in terms of the product of cost and throughput may be acceptably high. 

(b) Reduce the number of AWGs (and detector pixels)  by making each work in a partly multimode (i.e. few mode) configuration so that the AWG disperses light which is not in a single mode. This may be acceptable if high resolving power is not required e,g, in a survey of faint, unresolved galaxies.

(c) Reduce the number of AWGs (and detector pixels) by decreasing  the field of view. This directly trades-off cost with field coverage. This is of relevance to applications requiring little spatial multiplex, e.g. single-object spectroscopy or spectroscopy with high-order adaptive optics such that the input image is already near the diffraction limit \citep{Harris2012}.

\section{Conclusion}
\label{sec:conclusion}

We have examined the application of Integrated Photonic Spectrographs in astronomy and shown that an IPS is equivalent in function to an image slicer. We have shown that as the telescope diameter increases, the size of an IPS must also increase (provided that the slit is not matched to the diffraction-limit) due to the increase in number of modes in the field (equivalent to the number of diffraction-limited slices). 
We have also shown that the number of modes in a field is independent of how the field is initially sampled (the size of the sampling element (spaxel) has no effect on the total number of modes in the field).  \newline

We modelled IPS instruments to compare them with conventional instruments on large telescopes and found that they require 10$^3$-10$^5$ Arrayed Waveguide Gratings (AWGs) or 10$^{3}$-10$^{4}$ replica spectrographs if bulk optics are used for instruments on  8m and 30m telescopes. We found that fully-photonic instruments were comparable in size to their conventional counterparts but only if the AWG was close to the diffraction limit. The semi-photonic instruments were found to be much larger, due to the redundancies of having multiple spectrographs. \newline
We have also found that unless the input image is sampled near the diffraction limit, the number of component spectra in each spaxel is very high, requiring large numbers of pixels in the detector array. This is equivalent to oversampling the PSF and could also potentially increase detector noise in the instrument.  \newline

To combat the problem of size we considered the effect of adding extra inputs to the AWG to reduce the number of AWGs required. However the resulting instrument was of the same size or larger. It also means that the problem of oversampling in the linear case remains unsolved and potentially will be worse since the spectra will need to have gaps between them to distinguish them.

We also examined other options for reducing the number of detector pixels and/or AWGs and concluded that instruments of photonic construction may be viable depending on the extent to which performance (including throughput, spectral resolution and spatial multiplex)  can be traded against cost.

Even without these modifications or restrictions, there are some areas where IPSs may offer a significant advantage. These include spectroscopy of objects near the diffraction-limit, e.g. single objects with high-order AO such as in exoplanet studies. Another is low-resolution multiplexed spectroscopy working in the  few-mode limit.

\section*{Acknowledgements}
We gratefully acknowledge support from the Science and Technology Facilities council. We wish to thank the European Union for funding for the OPTICON Research Infrastructure for Optical/IR  astronomy under Framework Programme 7. We are very grateful for the AWG supplied by Gemfire UK and their technical expertise. We also wish to thank Robert Thomson for many useful discussions into all things Photonic and Tom Shanks for inadvertently asking a poignant question. We would finally like to thank everyone else who has taken the time to discuss the workings of various components with us, particularly Jon Lawrence, Nick Cvetojevic and Nemanja Jovanovic.

\bibliographystyle{mn2e}
\bibliography{ref}

\end{document}